\documentclass[12pt]{article}
\usepackage[mathscr]{eucal}
\usepackage{amssymb,latexsym}
\usepackage{verbatim}
\usepackage{amsmath}
\usepackage{amsthm}
\usepackage{enumerate}
\usepackage{authblk}
\usepackage{color}
\usepackage{url}

\newtheorem{thm}{Theorem}
\newtheorem{lem}[thm]{Lemma}
\newtheorem{prop}[thm]{Proposition}

\newcommand\bbR{{\mathbb R}}
\newcommand\bbZ{{\mathbb{Z}}}

\renewcommand\S{\Sigma}

\newcommand\D{\nabla}

\newcommand\ric{{\rm Ric}}

\newcommand\g{\gamma}

\renewcommand\th{\theta}

\newcommand\beq{\begin{equation}}
\newcommand\eeq{\end{equation}}
\newcommand\ben{\begin{enumerate}}
\newcommand\een{\end{enumerate}}
\newcommand\bit{\begin{itemize}}
\newcommand\eit{\end{itemize}}

\newcounter{mnotecount}

\title{Topology and singularities in cosmological spacetimes obeying the null energy condition}

\author{Gregory J. Galloway\thanks{Research partially supported  by NSF grant DMS-1313724.}
}
\author{Eric Ling}
\affil{Department of Mathematics
\\ University of Miami }

\begin{document}
\date{}
\maketitle

\begin{center}
{\it We dedicate this paper to the memory of Ted Frankel}
\end{center}
\vspace{.15in}

\begin{abstract}  We consider globally hyperbolic spacetimes with compact Cauchy surfaces in a setting compatible with the presence of a positive cosmological constant.  More specifically, for  $3+1$ dimensional spacetimes which satisfy the null energy condition and contain a future expanding compact Cauchy surface, we establish a precise connection between the topology of the Cauchy surfaces and the occurrence of past singularities.   In addition to the Penrose singularity theorem, the proof makes use of some recent advances in the topology of $3$-manifolds and of certain fundamental existence results for minimal surfaces. 

\end{abstract}


\section{Introduction}
\label{intro}

The classical Hawking cosmological singularity theorem \cite[p. 272]{HE} establishes  past timelike geodesic incompleteness in spatially closed spacetimes that at some stage are future expanding.  This singularity theorem requires the Ricci tensor of spacetime to satisfy the strong energy condition,  $\ric(X,X) \ge 0$ for all timelike vectors~$X$.  In spacetimes obeying the Einstein equations with positive cosmological constant,  $\Lambda > 0$, this energy condition is not in general satisfied, and the conclusion then need not hold; de Sitter space, which is geodesically complete, is an immediate example.  But this is not just a feature of vacuum spacetimes; dust filled FLRW spacetimes with positive cosmological constant provide other examples. For the FLRW models discussed in \cite[Section 3]{GBeem},
the co-moving Cauchy surfaces are assumed to be compact, and, apart from the time-dependent scale factor, have constant curvature $k = +1, 0, -1$.  These three cases are topologically quite distinct.  For instance, in the $k = +1$ (spherical space) case, the Cauchy surfaces have finite fundamental group, while in the $k = 0, -1$ (toroidal and hyperbolic $3$-manifold) cases, the fundamental group is infinite.   Moreover, it is only in the $k = +1$ case, that the past big-bang singularity can be avoided. 

In \cite{AndGal}, this topology dependent behavior was studied in a much broader context (not requiring any special symmetries) for spacetimes which are asymptotically de Sitter in the sense of admitting a regular spacelike conformal (Penrose) compactification.  Originally motivated by work of Witten and Yau \cite{WittenYau} pertaining to the AdS/CFT correspondence (see also \cite{Caigal}), the results obtained in \cite{AndGal} establish connections between the bulk spacetime (e.g., its being nonsingular) and the geometry and topology of the conformal boundary.  These results extend to this more general setting the behavior seen in the FLRW models.

In this note we present a result of a similar nature, which explicitly relates the occurrence of singularities in spacetime to the topology of its Cauchy surfaces.
 By taking advantage of advances in our understanding of the topology of $3$-manifolds, specfically the positive resolution of Thurston's geometrization conjecture, and subsequent consequences of it,  we are able to signficantly strengthen aspects of some of the results in \cite{AndGal} for $3+1$ dimensional spacetimes.  

The main aim is to prove the following.

\begin{thm}\label{main} 
Suppose $V$ is a smooth compact  spacelike Cauchy surface in a $3+1$  dimensional spacetime $(M,g)$  
that satisfies the null energy condition (NEC),
$\ric(X,X) \ge 0$ for all null vectors $X$.  Suppose further that $V$ is expanding in all directions (i.e. the second fundamental form of $V$ is positive definite; see section~\ref{proof} for details).   Then either
\ben
\item [(i)] V is a spherical space, or
\item [(ii)]  $M$ is past null geodesically incomplete. 
\een
\end{thm}

By a {\it spherical space}, we mean that $V$ is diffeomophic to a quotient of the 3-sphere $S^3$, $V = S^3/\Gamma$, where 
$\Gamma$ is isomorphic to a subgroup of $SO(4)$. Typical examples are the $3$-sphere itself, lens spaces and   the Poincar\'e dodecahedral space; for a complete list see e.g. \cite{Friedl} and references therein.  By taking quotients of de Sitter space, we see that there are geodesically complete spacetimes satisfying the assumptions of the theorem, having Cauchy surface topology that of any spherical space.  Nevertheless, one can view 
Theorem \ref{main}  as a  singularity theorem: Under the assumptions of the theorem, if $V$ is not a spherical space, i.e. if $V$ is not a $3$-sphere, or a quotient thereof, then $(M,g)$ is past null geodesically incomplete.  

The proof involves several geometrically interesting elements.  In addition to recent results in $3$-manifold topology, the proof makes use of a fundamental existence result for minimal surfaces due to  Schoen and Yau \cite{sysc}, in addition to a well known existence result coming from geometric measure theory.  Ultimately, the proof depends on (a slight refinement of) the Penrose singularity theorem.   
Finally, we would like to mention the paper \cite{Mukuno}, which played a role in motivating the present work.

\section{The proof}\label{proof}

We begin by presenting some preliminary results.  Let $(M,g)$ be a spacetime, i.e. a smooth time-oriented Lorentzian manifold.  We assume throughout that $M$ is $4$-dimensional.  Let $V$ be a smooth spacelike hypersurface in $M$ with induced metric $h$ and second fundamental form $K$.  To set sign conventions, put $K(X,Y) = g(\nabla_Xu,Y)$,
where $X,Y \in T_pV$, $\D$ is the Levi-Civita connection of $M$, and $u\in T_pM$ is the future directed timelike unit normal vector field to $V$.   Note that if $X$ is a unit tangent vector to $V$ which is extended locally by making it invariant under the normal geodesic flow then $K(X,X) = d/dt|_{t = 0}\|X\|$, where $t$ is proper time and $\|X\| = g(X,X)^{1/2}$.  Hence, we say that $V$ is \emph{expanding in all directions} if $K$ is positive definite.   As discussed in \cite{AndGal}, future asymptotically de Sitter spacetimes, in the sense of admitting a compact spacelike future conformal infinity, contain
Cauchy surfaces that are expanding in all directions.  We point out that in Hawking's singularity theorem it is the {\it trace} of $K$ (i.e. the mean curvature) that is  required to be positive.

We will need a slight refinement of the notion of a trapped surface.  Consider an immersion 
$f\colon\S \to V$, where $\S$ is a smooth compact surface; we refer to $\S$ as an immersed surface in 
$V$.  For each $p \in \S$, there exists a neighborhood $U \subset \S$ of $p$ such that $f|_U\colon U \to V$ is an embedding; hence $f|_U$ is a diffeomorphism onto its image $f(U)\subset V$.  Taking $U$ sufficiently small, there exists a smooth unit normal field $\nu_U$ along $f(U)$ in $V$; we refer to $U$ as an admissible neighborhood. $f(U)$ supports two past directed null normal vector fields $\ell_U^{\pm} = -u \pm \nu_U$.   Tracing the null second fundamental forms 
$\chi_U^{\pm}$ associated to the null normals $\ell_U^{\pm}$, one obtains the (past) null expansion scalars  
$\th_U^{\pm}$ along $f(U)$.   Note that (cf., \cite[Section 2.1]{AEM})
\beq\label{thid}
\th_U^{\pm} = -{\rm tr}_{f(U)} K \pm H_U  \,,
\eeq
where ${\rm tr}_{f(U)} K$ is the partial trace of $K$ with respect to the induced metric $\g_U$ on $f(U)$, and $H_U$ is the mean curvature  of $f(U)$ in $V$ with respect to $\nu_U$. 
Finally, $\S$ is said to be an {\it immersed past trapped} surface, provided there is 
a collection of admissible neighborhoods $U$ covering $\S$, such that $\th_U^+ < 0$ and $\th_U^- < 0$.

In the proof of Theorem \ref{main}, past trapped surfaces will arise as follows. An immersion  $f\colon \S \to V$ is said to be {\it minimal}, and $\S$ a minimal immersed surface, provided there is 
a collection of admissible neighborhoods $U$
covering $\S$ satisfying $H_U = 0$ for each $U$.  Note that this is independent of the collection of admissible neighborhoods covering $\S$.   Equation \ref{thid} then implies the following (see also \cite{GalJPhys}).

\begin{lem}\label{trapped}  
Suppose $V$ is a spacelike hypersurface in a spacetime $(M,g$) which is expanding in all directions.  If $\S$ is a compact minimal immersed surface in $V$, then $\S$ is past trapped. 
\end{lem}

The Penrose singularity theorem \cite{HE} remains valid for {\it immersed} trapped surfaces, with only minor modifications in the proof.   Thus we have the following.

\begin{thm}[Penrose]\label{penrose}
Let $(M,g)$ be a spacetime that satisfies the NEC and admits a smooth spacelike noncompact Cauchy surface $V$.  If $V$ contains a compact immersed past trapped surface $\S$, then $(M,g)$ is past null geodesically incomplete.
\end{thm}  

To apply (this version of) the Penrose singularity theorem to Theorem \ref{main}, one needs to come up with a noncompact Cauchy surface. This will be accomplished by finding a noncompact covering of the given Cauchy surface. Given a covering of a Cauchy surface, the next lemma gives a natural covering of the spacetime (see also \cite{EGP, Mukuno}).

\begin{lem}\label{cover} 
Let $V$ be a smooth spacelike Cauchy surface in a spacetime $(M,g)$ having induced metric $h$ and second fundamental form $K$.  Suppose $p\colon \tilde V \to V$ is a Riemannian covering, with metric $\tilde h =p^*h$ on $\tilde V$.  Then there exists a Lorentzian covering 
$P\colon \tilde M \to M$, with metric $\tilde g = P^*g$ on $\tilde M$, such that $\tilde V$ is a Cauchy surface for
$(\tilde M, \tilde g)$ with induced metric $\tilde h$ and second fundamental form $\tilde K = P^*K$.
\end{lem}

\proof One can use basic covering space theory to construct the desired spacetime $(\tilde M,\tilde g)$.  Alternatively, and somewhat more concretely, one can make use of a well known splitting result of Bernal and Sanchez \cite{Sanchez}.   By  \cite[Theorem 1.2]{Sanchez}, we may assume $M = \bbR \times V$, and 
$g = - \phi^2 dt^2 + h_t$, where $\phi$ is a smooth positive function on $ \mathbb{R} \times V$ and for each $t$,  $h_t$ is a Riemannian metric on $ \{t\} \times V$, with $h_0 = h$, so that we identify $V$ with $\{0\} \times V$.
Now consider the spacetime, $\tilde M =  \mathbb{R} \times \tilde V$, $\tilde g = - \tilde \phi^2 d t^2 + \tilde h_t$, where $\tilde \phi (t, x) = \phi \big(t, p(x)\big)$ and $\tilde h_t = p^* h_t$.  It is easily seen that $\tilde V = \{0\} \times \tilde V$ is a Cauchy surface for $(\tilde M, \tilde g)$.  Then $P : \tilde M \to M$, given by $P(t, x) = \big(t, p(x)\big)$, is the desired Lorentzian covering.\qed

\medskip  
The next proposition will be useful in the proof of Theorem \ref{main}.

\begin{prop}\label{2homology}  
Let $(M,g)$ be a $3+1$ dimensional globally hyperbolic spacetime that satisfies the NEC, and let $V$ be an orientable smooth compact spacelike Cauchy surface in $(M,g)$ that is expanding in all directions.  If $V$ has nontrivial second homology, $H_2(V,\bbZ) \ne 0$, then $M$ is past null geodesically incomplete.
\end{prop}

\proof The proof, which we sketch here, is essentially contained in the proof of Theorem 4.3 in \cite{AndGal}.  
By well known results of geometric measure theory (see \cite[p. 51]{Lawson} for discussion), every nontrivial class in $H_{2}(V,\mathbb Z)$ has a least area representative which can be expressed as  a linear combination of smooth, orientable, connected, compact, embedded minimal (mean curvature zero) surfaces in $V$.  Let $\Sigma$ be such a surface; we may assume  $\Sigma$ represents a nontrivial element of $H_{2}(V,\mathbb Z)$;  in particular, $\S$ is nonseparating. By Lemma \ref{trapped}, $\S$ is a past trapped surface.  Since $V$ is compact, we are not yet in a position to apply Theorem \ref{penrose}.  We must first pass to a suitable covering spacetime. 

Since $\S$ is nonseparating in $V$, there exists a loop $\g$ in $V$ that has intersection number $+1$ (with respect to appropriate orientations).   Using basic covering space techniques, one can pass to a Riemannian covering $p\colon \tilde V \to V$ which ``unravels" this loop (even if it is traversed infinitely often).  This covering can be described in terms of cut and paste operations as follows.   By making a cut along $\Sigma$, we obtain a compact manifold $W$ with two boundary components, each isometric to $\S$.  Taking $\mathbb Z$ copies of $W$, and gluing these copies end-to-end we obtain the noncompact covering manifold $\tilde V$ of $\S$, with covering map $p\colon \tilde V \to V$ defined in the obvious way.  In this covering, $\S$ is covered by $\mathbb Z$ copies of itself, each one separating $\tilde V$; let $\tilde \S_0$ be one such copy.   By Lemma \ref{cover} there exists a covering spacetime $P\colon \tilde M \to M$ such that $\tilde V$ is a Cauchy surface in $(\tilde M, \tilde g)$.  Moreover, by this lemma,  the `initial data' $h, K$ of $V$ lifts to $\tilde V$, i.e.,  $\tilde h = p^*h$ and $\tilde K = P^*K$. It follows that $\S_0$ is a past trapped surface in $(\tilde M, \tilde g)$.  Furthermore, since $P\colon \tilde M \to M$ is a local isometry, the null energy condition holds on $(\tilde M, \tilde g)$.  Proposition \ref{penrose} now implies that $(\tilde M, \tilde g)$ contains a past inextendible null geodesic $\tilde \eta$ that is past incomplete. It follows that the projection $\eta = P \circ \tilde \eta$ is past incomplete in $(M,g)$.\qed

\medskip
We are now ready to prove our main theorem. 

\proof[Proof of Theorem \ref{main}]   Assume that $M$ is past null geodesically complete.  We will show
that $V$ is a spherical space.  Suppose, at first, that $V$ is orientable.  Then by the prime decomposition theorem (see e.g. \cite{Hatcher}), together with the positive resolution of the elliptization conjecture (which includes the Poincar\'e conjecture, see e.g. \cite{Friedl} for discussion), $V$ can be expressed as a finite connected sum,
\beq\label{decomp}
V = V_1 \#V_2\# \cdots \# V_k \,,
\eeq 
where for each $i = 1,...k$, 
\ben
\item[(i)] $V_i$ is a spherical space, or
\item[(ii)] $V_i$ is diffeomorphic to $S^2 \times S^1$, or
\item[(iii)]  $V_i$ is a $K(\pi,1)$ manifold.
\een
Recall, a $K(\pi,1)$ manifold is a manifold, like the torus, whose universal cover is contractible;  in particular it has infinite fundamental group.  Both spherical spaces and $K(\pi,1)$'s are irreducible (i.e.  have the property that every embedded $2$-sphere bounds a ball).  

We first observe that there can be no $S^2 \times S^1$'s in the prime decomposition: If for some $i$, $V_i$ were diffeomorphic to $S^2 \times S^1$, $V$ would contain a nonseparating $2$-sphere, which would represent a nontrivial element of $H_2(V,\bbZ)$.  But then by Proposition \ref{2homology}, $(M,g)$ would be past null geodesically incomplete, contrary to assumption.  

We now show that there can be no $K(\pi,1)$'s in the the prime decomposition of $V$.  Suppose for some 
$i$, $V_i$ is a $K(\pi,1)$ manifold.  Then $V_i$ is irreducible, with infinite fundamental group.  Hence, by the positive resolution of the surface subgroup conjecture \cite{Kahn}, together with the 
fact that $\pi_1(V) = \pi_1(V_1) * \dotsb * \pi_1(V_k)$,
there is a subgroup $G$ of $\pi_1(V)$ isomorphic to the subgroup of a surface  genus $g \ge 1$.  Then by results in \cite{sysc}, there exists a minimal immersion 
$f\colon \S_g \to V$, where $\S_g$ is a surface of genus $g \ge 1$, such that the induced homomorphism on fundamental groups, $f_*\colon \pi_1(\S_g) \to \pi_1(V)$ is injective.  Basic existence theory for covering spaces guarantees the existence of a covering $p\colon\tilde V \to V$ such that $p_*(\pi_1(\tilde V)) = f_*\big(\pi_1(\S_g)\big)$. (It further follows that $p_*\colon\pi_1(\tilde V) \to f_*\big(\pi_1(\S_g)\big)$ is an isomorphism.)   Then by the standard map lifting criterion, there exists a map $F\colon \S_g \to \tilde V$ such that $p \circ F = f$.  Since $f$ is a minimal immersion and $p$ is a local isometry, it follows that $F$  is a minimal immersion.  Hence, $\S_g$ is a compact minimal immersed surface in $\tilde V$.

Now apply Lemma \ref{cover}:  There exists a spacetime $(\tilde M, \tilde g)$ such that $\tilde V$ is a Cauchy surface for $(\tilde M, \tilde g)$.  It further follows from this lemma and Lemma \ref{trapped} that 
$\S_g$ is a compact immersed past  trapped surface in $(\tilde M, \tilde g)$.  Moreover, since the fundamental group of $\tilde V$ is isomorphic to the fundamental group of a surface of genus 
$g \ge 1$, it must be noncompact (see for example \cite[Theorem 10.6]{Hempel}).   One can now apply Theorem \ref{penrose}  to conclude that $(\tilde M, \tilde g)$ is past null geodesically incomplete.  As in the proof of Proposition \ref{2homology}, this implies that $(M,g)$ is past null geodesically incomplete, contrary to assumption.

Thus, under the assumption of past null geodesic completeness we have now shown  that each $V_i$ in the connected sum \eqref{decomp} is a spherical space.  It remains to show that there is only one $V_i$, i.e. that $k =1$.   We now argue that if there were more than one (nontrivial) spherical space then there would be a finite covering $p\colon\tilde V \to V$, with $H_2(\tilde V, \bbZ) \ne 0$.  Then Lemma \ref{cover} and Proposition \ref{2homology} would imply that $(M,g)$ is past null geodesically incomplete.

Suppose 
\[
V = V_1 \#V_2\# \dotsb \# V_k
\]
where each $V_i$ is a spherical space, which, without loss of generality, we may assume is nontrivial.
We will use the following fact which is established by simple cut and paste arguments.
\
\
\
\begin{lem}\label{fact}
Let $X$ and $Y$ be closed $3$-manifolds.
Suppose $\tilde{X}$ is a finite cover of $X$ with $r > 1$ sheets, then 
\[
\tilde{X} \# \underbrace{Y \# \dotsb \#Y}_{r\: {\rm times}} \:\:\:\: \text{ is a cover of } \:\:\:\: X \#Y \,.
\]
\end{lem}
To continue the proof, let $W = V_3 \# \dotsb \#V_k$ so that $V = V_1 \# V_2 \# W$. Since, for closed 3-manifolds $M$ and $N$,  $H_2(M \# N) = H_2(M)\oplus H_2(N)$, Lemma \ref{fact} implies that it suffices to show $V_1 \# V_2$ admits a finite sheeted cover $V'$ with $H_2(V') \neq 0$.

To accomplish this consider the universal cover of $V_1$ which is just $S^3$. $S^3$ covers $V_1$ with $r > 1$ sheets. Lemma \ref{fact} implies 
\[
S^3 \# V_2 \# \dotsb \#V_2 \approx V_2 \# \dotsb \#V_2 \:\:\: \text{is a finite cover of } V_1 \# V_2.
\]
Another application of Lemma 6 implies that it suffices to show $V_2 \# V_2$ admits a finite sheeted cover $V''$ with $H_2(V'') \neq 0$. 

This may be achieved as follows. By removing a suitable handle $H \approx I \times S^2$ from 
$V_2 \# V_2$, we obtain two copies of $V_2 \setminus B$, where $B$ is an open ball.  The univeral cover of each is a $3$-sphere minus $s > 1$ copies of $B$. By appropriately gluing back in $s$ copies of the handle 
$H$, we obtain a covering manifold $V''$ consisting of two $3$-spheres connected by $s$ handles.  Since $s > 1$ we see that $H_2(V'') \neq 0$.

Hence, under the assumption that  $(M,g)$ is past null geodesically complete, we have shown that $V$ is a spherical  space.  This completes the proof of Theorem \ref{main} in the case that $V$ is orientable.  
Now suppose that $V$ is nonorientable.  Let $\tilde V$ be the orientable double cover of $V$, and apply Theorem \ref{main} to obtain the covering spacetime $(\tilde M, \tilde g)$ with Cauchy surface $\tilde V$.   Hence either $\tilde V$ is a spherical space or $(\tilde M, \tilde g)$ is past null geodesically incomplete.  If $\tilde V$ is a spherical space then $\tilde V$, and hence $V$, are covered by the $3$-sphere.  This implies that 
$V$ has finite fundamental group.  But this contradicts the fact that a closed nonorientable $3$-manifold must have infinite fundamental group (see e.g. \cite[Lemma 6.7]{Hempel}).   Thus, $(\tilde M, \tilde g)$, and hence 
$(M,g)$,  are past null geodesically incomplete.  This completes the proof of Theorem \ref{main}.\qed

\bibliographystyle{amsplain}
\bibliography{cosmo}

\end{document}